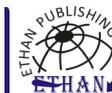

# Phase Transformation in Self-Organized Carbon Tribolayers

Manuel Mailian[1] and Aram Mailian[2]

1. LTX-Credence Armenia, 2 Adonts str., Yerevan 0014, Armenia

2. Institute for Informatics, 1 P. Sevak str., Yerevan 0014, Armenia

Corresponding author: Aram Mailian (amailian@ipia.sci.am)

**Abstract:** The simplest way to obtain thin carbon layers is to draw or rub with a graphite rod. During rubbing, forces of friction acting in graphite/substrate tribological system cause drastic changes in the structure of the interface stratum developing thereby stable self-organized and ordered thin structure. We present a pioneering experimental investigation of structural and morphological transformations in carbon tribolayers (CTL). By optical microscopy observation it is found that CTL is a multilayer structure, the essential building block of which is a transparent phase shaped as a lamina in-between the surface and bottom disordered layers of CTL. The surface of the lamina exhibits non-linear electrical conductivity near zero bias on I-V characteristics. The optical properties of the whole CTL are mostly controlled by physical processes occurring in the transparent lamina. The Raman spectrum of CTL contains narrow bands at 1,589 nm$^{-1}$ and 1,346 nm$^{-1}$ corresponding to G and D bands of carbon crystal lattice. The observed features are interpreted using the relationship between the bond length and corresponding band frequency, $r^2\omega = const$. Optical absorption of CTL has a feature at 4.6 eV originating from strong electron-hole interaction. From comparative analysis of experimentally data, structural-spectral correspondence is found. It is concluded that because of phase transformation during rubbing, a carbon structure consisting of sp$^3$ lamina with a nano-scaled thick sp$^2$ layer on the top is shaped.

## 1. Introduction

In recent years, intrinsic sliced morphology of hexagonal sp$^2$ bonded graphite has ignited a substantial interest toward obtaining and exploration of nanoscale thick layered carbon structures. Typically, the fabrication of such structures requires high-tech equipment and expensive materials. Therefore, as might be expected, the search of simple methods constantly was and currently is on the agenda of applied research. Meanwhile, the simplest way of fabrication of thin carbon layers, namely rubbing bulk graphite against a substrate, attracted researchers' attention just recently [1-7]. The main conclusion that drawn from these works is that rubbed layers behave like solid state thin films; therefore they can be used as sensors. Their physical properties and structural features, however, are largely unstudied up to now, hence publications on the subject are rare. The reason is hidden, first of all, in underrating the layers' structural performance and hence their unsteady and unpredictable physical characteristics.

Basically, the origin and development of rubbed carbon layers occur in the tribological system of graphite/substrate, hence structural changes and layer modification occur in the strata contiguous to the friction interface. The only creating source here is the force of friction. Actually, on nanoscale scene a permanent and repetitive process of distortion, breaking and re-building of chemical bonds take place under influence of forces of friction. This process results in a self-organized, dissipative layer which appears as a nanoscale thick and crystallographically ordered separate phase on the surface of the whole structure [8].

In case of graphite, the energy of friction can break or distort, first of all, weaker π bonds between hexagonal $sp^2$ bonded planes. Since this bond is responsible for electrical conductivity and optical properties of graphite, the process will naturally lead to an altered structure and hence to the altered parameters of rubbed off layers. Also, new carbon allotropes built of distorted $sp^2$ and $sp^3$ sites are probable to emerge spontaneously during friction.

Hence, there is significant motivation to suppose that it is possible to obtain a material of rubbing outcome with predictable and reproducible characteristics, needed so much for practical application. We believe also, that due to its nanoscale thickness and flexibility CTL is a promising material for thin-film and flexible electronics. It is natural to expect also, that as an ordered crystal film it should exhibit attractive optical characteristics.

The hope of perspective application requires detailed study of these layers' structure and morphology as well as physical, optical and electrical properties. We present the pioneering study focusing on transformations in CTL structure and morphology.

## 2. Experimental

The procedure of obtaining of a CTL included the following steps. First, the substrate was prepared by abrading the surface of crystalline NaCl bar with sandpaper of 30 m abrasive coating along y-axis (Fig. 1a). Obtained rough surface contained straight and parallel scratches dug by abrasive grains. Then, the graphite layer was obtained by carefully and repeatedly rubbing a rod of graphite along x-axis (at 90° to the substrate surface) against the prepared surface (Fig. 1b). The pressure of 5 MP applied on the rod was kept constant during the whole rubbing cycle.

To monitor CTL evolution, the graphite trace was observed under an optical microscope both in transmission and absorption modes. The angle of side illumination in reflection mode was taken close to 20° to the surface plane.

Freestanding CTLs were separated by employing a simple technique [9, 10]. The NaCl crystals with CTLs on the surface were immersed into water and after the dissolution of substrate, the detached structure released and floated in the water (Figs. 1d). Subsequently, the structure was transferred onto a glass substrate picking it up carefully (Fig. 1f). This technique allows transferring the CTLs onto the substrate in two positions: rubbed face up or down.

Electrical measurements were performed using the two-probe method by gently pressing gold or copper contacts on the structure surface.

With experimental procedure, we observed no qualitative but numerical difference in behavior and physical properties of layers obtained by rubbing with different commercially available graphite rods.

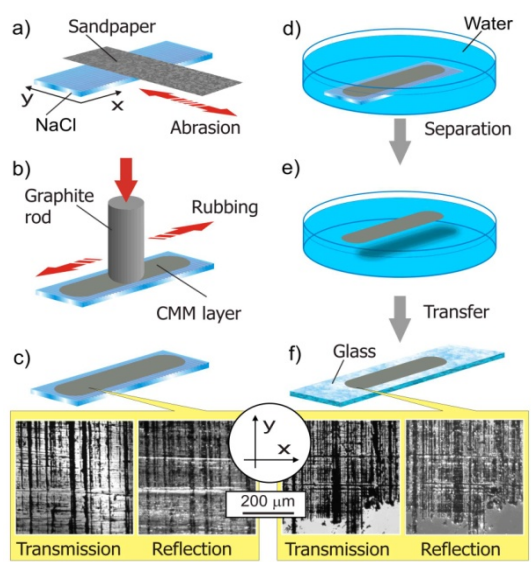

Fig. 1  Obtaining and separation of CTL. a) Processing of the surface of NaCl substrate; b) Obtaining of graphite layers by rubbing; c) Schematic presentation and microscopic images of an area of CTL on NaCl substrate in transmission and reflection modes of illumination; d, e) Separation of obtained layers; f) Microscopic mages of the same area of transferred layer in transmission and reflection modes of illumination. The CTL is transferred rubbed face up. Round inset is given to mention the direction of dark strips.

## 3. Results and Discussion

Before discussing experimental results, it is useful to introduce the specific feature of rubbing process. The initial stroke of graphite rod occurs between the parent graphite bulk and bare substrate. Then, with each stroke, the carbon layer evolves and the friction occurs between the surface layer of the graphite bulk and the surface of the CTL modified during previous stroke. This dynamic process of layer development suggests that physical processes should take place at perpetual changing interface between the friction surfaces of rubbed layer and bulk graphite. From tribological point of view, during rubbing the layers are formed and modified by a combined mechanical action of sliding and cleaving the basic layers close to surface of graphite bulk, transferring and pressing on the substrate surface the graphite flakes. The tribological characteristics of the graphite/substrate system, wear rate at friction interface and the coefficient of friction vary monotonically with the stroke number. We observed that they reach the stable low value after a certain number of strokes. We studied CTL with stabilized tribological characteristics.

The main feature found by the optical observation of CTLs is that large areas between darker strips remain visually transparent during the whole cycle of rubbing (Fig. 1c, left callout). The estimation of optical transmission of randomly chosen small transparent areas reveals that the layer transparency can reach 95%.

In closer inspection, two series of dark strips in the texture of the optical image of rubbed layers are seen in transmission mode of observation. The first series lies along scratches on the substrate surface and reflects the features of the scratches and their appearance does not change much during the whole cycle of rubbing. Obviously, dark strips are the projection of thicker bars formed by the filling the scratches with graphite flakes. The second series lies along the rubbing path and apparently on the rubbed face. The width, darkness and location of these strips vary with each stroke. Because we rubbed the rod perpendicular to scratches, dark strips formed check-like pattern in transmittance projection (Fig. 1c).

The dark strips along x-axis produced by rubbing turn into partly shining ones, while the others (along y-axis) fade when side illumination is applied (Fig. 1c, right callout). Evidently, this effect is caused by light reflection from the slopes of thicker ridge-like structures projected as dark strips at the surface.

The same texture and effect of shining is observed for transferred layers (Fig. 1f, callouts). In sketch, the CTL consists of three structural elements (Fig. 2): graphite bars formed by filling scratches with graphite grains, the transparent lamina above the bars and ridge-like structures on the top of the structure [9].

The existence of a transparent phase within a CTL can be interpreted in either of two ways: (1) the transparent lamina should be thin enough (several basic planes of carbon, i. e. few-layer graphene), highly transparent to light, or (2) during rubbing, the morphology of rubbed layer should have changed so much that turn the thick layer into transparent one. We ruled out the existence of bare $sp^2$ few-layer graphene since Atomic Force Microscopy observation of the transparent lamina revealed no area thinner than 20 nm which should not be visually transparent since a monolayer of $sp^2$ carbon absorbs 2.3% of incident light in visible light region.

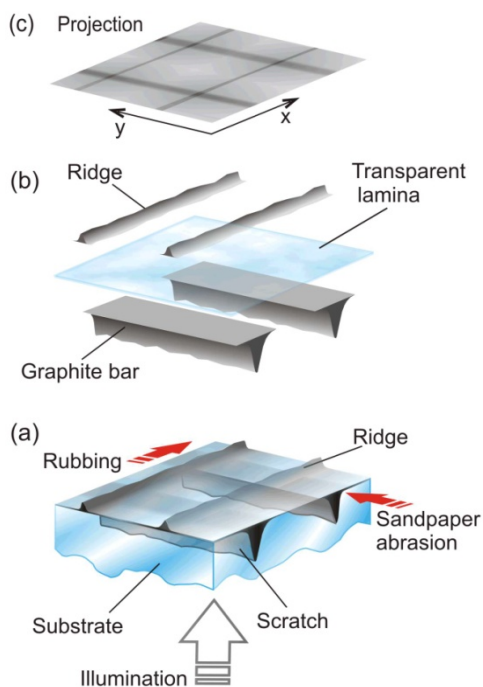

Fig. 2  Schematic presentation of CTL structure. a) A segment of the CRL on the NaCl substrate; b) The building blocks of the same segment; c) The segment's projection on microscope objective in transmission mode.

On the other hand, studies of the electrical behavior of the same transparent area hint at the presence of a nanoscale thick layer on the surface of the transparent lamina. Namely, when the measuring contacts are applied onto thicker and higher conductive graphite bars emerging on the surface of a transparent segment (Figs. 3a, b), I-V measurements reveal non-linear behavior at the vicinity of zero bias (Fig. 3d). The latter is characteristic of a carbon structure containing few layers of graphene.

In addition, other observations-electrical and temperature behavior of electrical conductivity indicate that the transparent lamina is a crystalline insulator or a high-resistant semiconductor [9].

Transparent areas allowed us to carry out optical spectroscopy measurements. The optical absorption spectrum was obtained over the range of photon energies 1.2-6.2 eV (Fig. 4). Its behavior is very similar to that of graphene [11-13]. The absorption spectrum displays a couple of noteworthy features. First, a pronounced resonance arises in the ultraviolet at 4.6 eV. This transition takes place at M saddle-point Van Hove singularity of the Brillouin zone of graphene [11-13]. The observed resonance feature has an asymmetric line shape, with sharp increase in absorption on higher photon energy site.

Second, the absorbance increases beginning from ~5.5 eV. This latter is not reported for graphene or graphite in the literature which indicates that the mechanism beyond the effect is unique to CTL.

GW calculations, which describe the bands of graphene the best, predict a band-to-band transition energy of $E_{GW}$ = 5.2 eV at the M-point of Brillouin zone if no interaction between electrons and holes is taken into account (Fig. 4a). Optical transition in CTL is red shifted nearly 0.6 eV from that value. This peculiarity is explained by the dominant role of electron-hole couples (excitons) in optical transitions [11-13].

Now, returning to the quality dualism - thick transparent lamina with the non-linear electrical behavior at the surface and optical response characteristic to nanoscale thick graphene structure-there is an easy explanation if we use Raman spectroscopy. Since the transparent areas occupy the considerable part of a CTL, one can expect that the Raman scattering signal from that area will dominate in the overall signal coming from the CTL surface.

The Raman spectra of a CTL contain two first order quite sharp modes reflecting the morphology of carbon allotropes (Fig. 4b). (1) G-band reflects bond vibration in pure $sp^2$ carbon structures (at 1,589 $cm^{-1}$ in our case); (2) D-band (at 1,346 $cm^{-1}$ in our case) supposed to origin from vibration of bonds of structural defects or bond deviations. In this study we applied a method of comparison of band frequencies to extract the length of bonds responsible for a given feature.

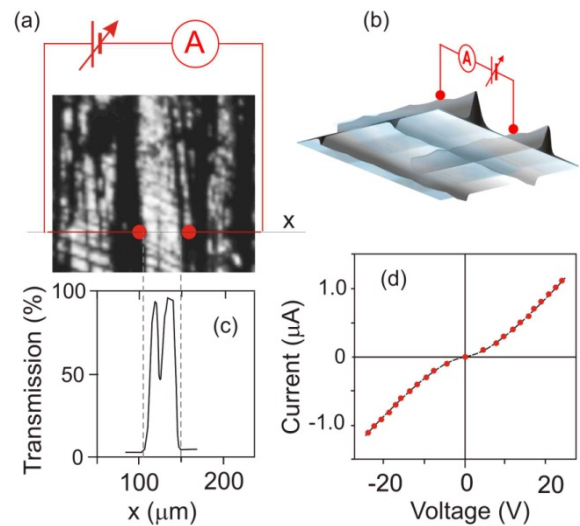

**Fig. 3** Visual presentation and characteristics of a segment of transferred CTL containing a transparent area between dark strips. a) Microscopy image of the segment and I-V measurement diagram. CTL is positioned rubbed face down; b) 3D schematic of I-V measurement diagram; c) Optical transmission along the straight line connecting the contacts; d) I-V characteristics.

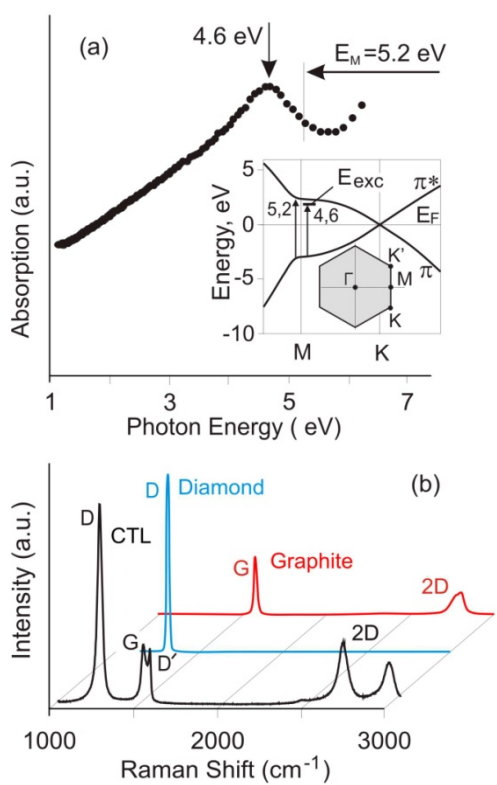

Fig. 4  a) The optical absorption spectrum of CTL. At right bottom the band diagram and Briilouin zone are shown. $E_M$ is assigned for interband optical transition at M point of Brillouin zone and $E_{exc}$- for exciton energy; b) The Raman spectrum of a CTL obtained on the surface of NaCl crystal at excitation wavelength of 514 nm. For comparison, Raman spectra of diamond (blue) and graphite (red) are shown too.

Upon further analysis our methodological approach for characterizing the bonding configuration of CTL must be explained. Both $sp^3$ bond and disordered $sp^2$ give rise to bands around 1,340 cm$^{-1}$ in Raman spectra. Very often the bands that originate from coexisting hexagonal $sp^3$ and imperfect $sp^2$ phases overlap and broader and flatter line shape is observed. In order to tell the fingerprints of structural deviations apart, one has to estimate the bond length by means of vibration frequencies of different C-C bonds responsible for D band. Since the most stable state is $sp^2$ hexagonal carbon structure with the shortest bond, in our approach we meant that any deviation from hexagonal $sp^2$ structure can be considered as a disordered or defected structure yielding D band. Then, pure tetrahedral diamond with longer $sp^3$ site could be considered as the furthermost stable deviation from $sp^2$ structure. Therefore, only $sp^3$ bonded diamond yields Raman spectra with a single narrow D band, whereas only $sp^2$ bonded pure graphite or graphene crystals have no D band. Considering these two stable carbon allotropes as a model states we will be able characterize the structural features of other allotropes.

Diamond-like-Carbon (DLC), a structure in-between hexagonal graphene and tetrahedral diamond, is built of differently distorted and deformed sites of $sp^3$ and $sp^2$ bonds [14]. This makes the Raman spectra of DLC flat with undistinguished broad bands. Narrower is the band, more ordered and perfect is the structure responsible for the band [14]. The Raman spectrum of a CTL contains both narrow G and D bands (Fig. 4b). The FHWM (Full Width at Half Maximum) of G and D bands are 25 and 30 cm$^{-1}$ respectively which speaks out about the existence of highly ordered crystalline structures of $sp^2$ and $sp^3$ phases in a CTL.

Early studies of stretching vibration of covalent bonds [15, 16] have disclosed that the stretching frequency of diatom bonds relate to the bond length as:

$$r^2\omega = const \qquad (1)$$

Later this rule was elucidated using a free-electron interpretation for a series of diatomic molecules [17]. For C-C bond they obtained the constant value 2.8 × 10$^{13}$ cm over all electronic states. For the Raman spectrum, shown in Fig. 4b, we obtained 3.2 × 10$^{13}$ cm for both G and D bands.

Eq. (1) can provide a link between the frequencies of G and D basic modes and their morphology in covalent bonded carbon allotropes. As follows from Eq. (1), graphene ($sp^2$ hybridization) and Diamond ($sp^3$ hybridization), relate to the bond length through a constant for a given excitation wavelength:

$$\left(\frac{r_D}{r_G}\right)^2 = \frac{\omega_G}{\omega_D} = const \qquad (2)$$

where: $r_G$ and $r_D$ are the lengths of $sp^2$ and $sp^3$ covalent bonds in graphene and diamond respectively, $\omega_D$ and $\omega_G$ are frequencies of D and G modes in Raman spectra of diamond and graphene respectively. We used Eq. (2) for revealing structural-spectral correspondence for CTLs. First, with $r_G = 142$ pm (pure graphite), $r_D = 154$ pm (pure diamond), and correspondingly $\omega_D = 1,332$ cm$^{-1}$ and $\omega_G = 1,580$ cm$^{-1}$ [18], we determined the constant equal to about 1.18 for the excitation wavelength of 514 nm. Then, applying the same formula and having $\omega_G = 1,589$ cm$^{-1}$ for the CTL we obtained ~154 pm for the bond length responsible for D peak at 1,346 cm$^{-1}$. Hence, we concluded that in a CTL the transparent phase is mainly $sp^3$ hybridized DLC which yields a characteristic D band.

Combined analysis of the Raman spectrum and the Fano theory of interaction between a separate exciton state and the conduction band continuum, reveals the mechanism behind the optical absorption beyond 5.5 eV. A natural way to study the effect is to examine the absorption of the whole transparent layer. As it is known the energy gap between π and π* bands is about 5.5 eV in diamond or DLC structures and light absorption exhibits sharp increase beyond that energy. It makes these materials transparent to visible light. Applied to our results, this means that during rubbing the transformation of a higher conductive and opaque $sp^2$ phase to a transparent and high-resistant $sp^3$ phase occurs. The effect of the formation of a self-organized DLC phase due to friction is recognized in tribological studies of graphite [19].

The Raman spectrum has two more essential features: bands are narrow and the peak intensity ratio $I_{2D}/I_G$ is about 1. They explain the coexistence of prominent D and G and 2D bands and reveal their correspondence to CTL structure and morphology.

The first feature indicates on presence of highly ordered crystalline structures. D band reflecting disorder should be shaped by overlapping the signals coming from the vibration of $sp^3$ 154 pm long and distorted $sp^2$ bonds in DLC. The bandwidth gives the bond length deviation about ± 2 pm, which advocates to high performance of the structure responsible for the band.

The second feature shows that CTL contains a structure composed of two or three graphene layers ($sp^2$) [18]. We rule out the origin of the Raman signal from scattering in clusters of different morphology, since in that case the Raman bands should be flat and broad [14]. Inasmuch as the electrical behavior of the surface of transparent area of CTL is non-linear (Fig. 3d) which is characteristic of nanoscale thin carbon layers and since the Raman spectrum unavoidably characterizes the surface, we claim that reversion from $sp^3$ to more stable $sp^2$ phase takes place within nanoscale thick stratum on the surface of thicker DLC [20, 21]. The layers of thicker DLC and few-layer graphene have different band gaps and therefore should have different optical characteristics. Consequently, the full optical spectrum of sandwiched CTL should be the resultant of the overlapping of two characteristics. For this reason, the $sp^2$ bonded graphene structure will dominate in optical absorption bellow 5.5 eV exhibiting a prominent peak at around 4.6 eV, while DLC will shape the spectrum mainly beyond that energy with sharp increase in absorption.

## 4. Conclusions

Rubbing of graphite bulk against a substrate yields a CTL structure composed of a transparent $sp^3$ bonded DLC lamina, sandwiched with a few-layer graphene on top. Due to self-organization, in CTL "dry reactions" of $sp^2$-$sp^3$ conversion and $sp^3$-$sp^2$ at the layer surface occur during rubbing. Ordered crystal phases in CTL gives rise to Raman spectra containing characteristic bands for both $sp^2$ and $sp^3$ carbon allotropes.

## Acknowledgment

The authors are grateful to Dr. Narine Ghazaryan, Orbeli Institute of Physiology, Yerevan, Armenia, for her help in obtaining optical spectra.